# Phononics of Graphene and Graphene Composites

**Alexander A. Balandin**

Phonon Optimized Engineered Materials (POEM) Center
Department of Electrical and Computer Engineering
Materials Science and Engineering Program
University of California – Riverside
Riverside, California 92521 U.S.A.
balandin@ece.ucr.edu

**Abstract:** I present a concise account concerning the emergence of a research field, which deals with the thermal properties of graphene, covering the refinement of understanding of phonon transport in two-dimensional material systems. The practical application of graphene and few-layer graphene in thermal interface materials are also discussed.

## I. Introduction

The field of phononics is comprised of the study of quanta of crystal lattice vibrations whose characteristics influence elastic, acoustic, and thermal properties of bulk and nanostructured materials[1-3]. Phononics of lower-dimensional material systems is particularly interesting, allowing one to elucidate the physics of crystal lattice vibrations and engineer the phonon spectrum to achieve new functionalities of the materials. Graphene – a monoatomic plane of $sp^2$-hybridized carbon atoms – is an excellent example of a material system of low dimensionality. The initial interest in graphene originated from its unique *linear* energy dispersion for electrons revealed in an unusually high charge carrier mobility, and other exotic electronic and optical properties[4-8]. Electrons are not the only elemental excitations influenced by a reduction in dimensionality which behave very differently from their counterparts in three-dimensional (3-D) bulk crystals. Phonons – both optical and acoustic – also demonstrate a significant sensitivity to the number of atomic planes as the sample thickness approaches the single-atomic-plane limit. It is no coincidence that monitoring the Raman spectral signatures of graphene's optical phonons – *G* peak and *2D* band – became a standard technique for the identification of graphene, and for counting the number of atomic planes in few-layer graphene (FLG)[9-11]. The adoption of Raman spectroscopy of graphene for this purpose became instrumental in the proliferation of graphene research: it is much easier to take a Raman spectrum than to conduct low-temperature transport measurements to identify single layer graphene, as was done previously.

Acoustic phonons, which are the dominant heat carriers in many materials, demonstrate a similar sensitivity to the number of atomic planes in FLG, changing their ability to conduct heat[12-14]. Graphene is not exactly a *true* two-dimensional (2-D) system for phonons, owing to the out-of-plane atomic motion, but it is as close as one can get to a 2-D system for phonons in the physical world [13-15]. The absence of inter-atomic-plane coupling in graphene and a modified phonon density of states leads to exotic thermal conductivity characteristics such as exceptionally high values and the dependence of the *intrinsic* thermal conductivity on the lateral size of the samples[14]. The availability of FLG allowed for the study of the evolution of thermal conductivity in thin films, which is limited by the intrinsic phonon dynamics rather than by the *extrinsic* effects, such as phonon scattering on rough interfaces[13]. In addition, FLG has made it possible to engineer the phonon dispersion in the entire Brillouin zone (BZ), and in the energy range, from acoustic to optical phonons, by a simple twist of one atomic plane about its surface normal vector with respect to another[16-20]. Apart from the intriguing fundamental science, the discovery of unique phonon transport characteristics of graphene motivated numerous studies of practical use of graphene and FLG in composites and coatings[21-26].

## II. Thermal Conductivity of Graphene and Few-Layer Graphene

The first experimental measurements of thermal conductivity of graphene were performed using a non-contact Raman optothermal method[12-15]. The Raman *G* peak of graphene is narrow, and its position is sensitive to temperature[27-28]. These attributes of the *G* peak allow one to use the peak's calibrated spectral position for determining a local temperature of the sample. The atomic thickness of graphene, which limits the heat flux, opens an opportunity for using a Raman spectrometer, with a conventional low-power laser, for measuring the highly conductive crystalline materials. The suspension of the graphene sample is another essential consideration of this method, necessary for the determination of the power dissipated in the graphene and ensuring the heat flux propagation along a graphene layer toward heat sinks[12-15, 29-32] (see Figure 1). Since its introduction, the Raman optothermal technique has been extended to a range of other 2-D materials beyond graphene[33].

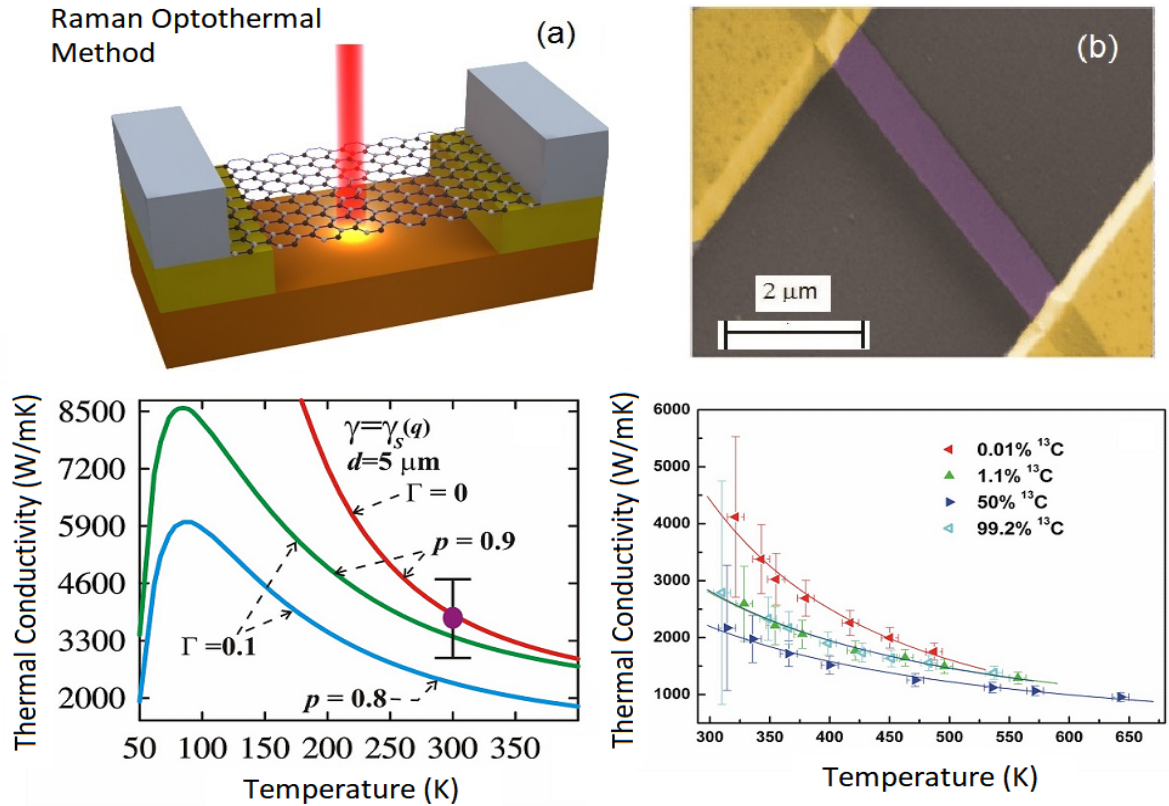

**Figure 1** (a) Schematic of the micro-Raman optothermal measurements. (b) Scanning electron microscopy image of a graphene ribbon suspended across a trench in $Si/SiO_2$ wafer. (c) Calculated thermal conductivity of graphene as a function of temperature for a graphene ribbon with the width of 5 μm. The results are shown for two values of the specularity parameter $p$ and point-defect scattering strength $\Gamma$. An experimental data point is provided for comparison. (d) Measured thermal conductivity of suspended CVD graphene with different concentration of $^{13}C$ isotope. The data are from Refs. [12, 14, 32]. The figures are reprinted with permission from the Nature Publishing Group.

The values of thermal conductivity of high-quality exfoliated *suspended* graphene measured by different groups using the Raman optothermal methods range from ~2000 W/mK to ~5000 W/mK near room temperature (RT)[14-15, 29-32]. The average measured thermal conductivity for high-quality exfoliated graphene is ~3000 – 4000 W/mK while that of high-quality chemical vapor deposition (CVD) polycrystalline graphene is ~2500 W/mK. The electrical measurements of thermal conductivity of CVD graphene, *e.g.* by the thermal bridge method, revealed ~2500 W/mK near RT[34-35]. The thermal conductivity of graphene supported on a substrate is reduced due to the coupling of graphene phonon modes to the substrate and additional phonon scattering on the graphene—substrate interface[36]. The thermal bridge measurements are considered to be more accurate but they suffer from experimental uncertainty related to unavoidable sample damage and contaminations during the required nanofabrication of the heaters and sensors[14]. Defects in the samples lead to lower thermal conductivity. The range of measured values can be attributed to (i) fundamental size dependence of the thermal conductivity of graphene owing to its 2-D nature; (ii) differences in the sample quality and geometry; (iii) limited accuracy of all experimental techniques; and (iv) mechanical strain and stress in the suspended samples. One should note that the lower bound of the thermal conductivity of graphene, exceeding that of basal planes of bulk graphite, is more important than the upper bound. The experimental thermal conductivity of carbon nanotubes has been reported in a similar range of values from ~1760 W/mK to 5800 W/mK[37]. The average conductivities for carbon nanotubes are at 3000 – 3500 W/mK[38-39]. The theoretical reports for graphene give a larger range of the thermal conductivity – from ~1000 W/mK to 10000 W/mK[15]. However, the theoretical and computational studies are in agreement that the intrinsic thermal conductivity of graphene should be larger than that of the carbon nanotubes[15, 40-44].

The thermal conductivity of suspended CVD graphene has been investigated as a function of the density of crystal lattice defects, introduced by the low-energy electron beam irradiation. The near-RT thermal conductivity decreases from ~ 2000 W/mK to ~ 400 W/mK as the defect density increases from $2.0\times10^{10}$ cm$^{-2}$ to $1.8\times10^{11}$ cm$^{-2}$. The thermal conductivity reveals an intriguing saturation behavior at higher concentration of defects[45]. The optothermal Raman technique has been used to investigate the phonon transport in the twisted bilayer graphene (TBG)[46]. In a wide range of examined temperatures, from 300 K to 750 K, the thermal conductivity in twisted bilayer graphene is smaller than both in graphene and naturally AB-stacked bilayer graphene (BLG). The thermal conductivity of TBG is by a factor of two smaller than that in graphene and by a factor of ~1.35 smaller than that in AB-stacked bilayer graphene near RT. The drop in thermal conductivity is explained by the emergence of many hybrid-folded phonons in TBG, resulting in more intensive phonon scattering[46]. The possibility of tuning the thermal conductivity of graphene by *isotope engineering* has also been demonstrated (see Figure 1).

## III. Phonons in Graphene and Bilayer Graphene

Investigation of heat conduction in graphene raised the issue of ambiguity in the definition of intrinsic thermal conductivity for 2-D crystal lattices. The thermal conductivity, *K*, limited by the crystal anharmonicity alone, referred to as *intrinsic*, has a finite value in 3-D bulk crystals[47]. However, the intrinsic thermal conductivity reveals a logarithmic divergence in 2-D crystals, $K \sim \ln(L)$, with the system size, *L*. This anomalous behavior, which leads to infinite thermal conductivity in 2-D systems, is different from the *ballistic* heat conduction in structures smaller in size than the phonon mean-free path[14]. The logarithmic divergence is related to the dimensionality and corresponding phonon density of states. Graphene is not a true 2-D system since it allows for the out-of-plane vibrations of the atoms, *e.g.* associated with the ZA and ZO phonon branches. As a result, the size dependence in graphene may deviate from the logarithmic, and the finite intrinsic limit can be achieved at some size of the sample. A recent theoretical work suggests that the convergence to the intrinsic thermal conductivity requires the sample size as large as a millimeter (mm)[43]. At such a length scale, in real graphene samples, the extrinsic scattering mechanism, *e.g.* phonon scattering on defects or grain boundaries, would unavoidably start inhibiting the heat conduction. Thus, the actual upper bound of the intrinsic thermal conductivity of graphene may remain an intellectual curiosity rather than a well-defined, observable quantity. The lower bound value is more relevant. If it is higher than that of a basal plane of graphite, then one can talk about the specific thermal conductivity of graphene which demonstrates the unique size dependence related to its 2-D nature. In other words, 2-D nature of the phonon density of states in graphene results in exceptionally long phonon mean free path (MFP) for the long-wavelength acoustic phonons, and corresponding high thermal conductivity. A number of theoretical and computational studies support this conclusion. A recent experimental study suggested logarithmic size dependence in graphene samples, differing from the linear dependence associated with the ballistic transport regime[48].

Graphene reveals four graphene sheet in-plane phonon branches: transverse acoustic (TA), longitudinal acoustic (LA), transverse optical (TO) and longitudinal optical (LO); and two out-of-plane acoustic (ZA) and optical (ZO) branches with the displacements perpendicular to the graphene plane. The in-plane acoustic branches are characterized by the linear energy dispersions over most of the Brillion zone (BZ) except near the zone edge while the out-of-plane ZA branch demonstrates a quadratic dispersion near the zone center $q = 0$, where *q* is the phonon wavenumber. The number of phonon branches in bilayer graphene is doubled: six additional branches possess non-zero frequency at $q = 0$ and at low frequencies they are affected by inter-layer interactions. The emergence of many folded hybrid phonon branches in TBG is explained by the change of the unit cell size and a corresponding modification of the reciprocal space geometry. The number of polarization branches and their dispersion in TBG depend strongly on the rotation angle (see Figure 2). TBG and FLG present interesting material systems where phonon dispersion can be engineered over the entire BZ and range of energies, from acoustic to optical phonons, by rotating atomic plane with respect to each other. Theoretical studies suggested that ZA phonons are the primary carriers that determine the specific heat for T ≤ 200 K while contributions from both in-plane and out-of-plane acoustic phonons are dominant for 200 K ≤ T ≤ 500 K. In the high-temperature limit, T > 1000 K, the optical and acoustic phonons contribute approximately equally to the specific heat[19-20]. The Debye temperature for graphene and twisted bilayer graphene is calculated to be around ~1861–1864 K. One can envision the possibility of engineering the thermodynamic properties of materials such as bilayer graphene, at the atomic scale, by controlled rotation of the $sp^2$-carbon planes. Another interesting question in the physics of phonons in graphene is the relative contribution of ZA phonons to the thermal conductivity. The original models neglected or underestimated ZA phonon contributions in consideration of their low phonon group velocity and large anharmonicity, characterized by the large Grüneisen parameter. Later, it has been suggested theoretically that due to phonon-scattering selection rules, ZA phonons are long-lived and can account for a significant portion of overall heat conduction, at least, in certain temperature ranges[40-41, 49]. However, it has been also pointed out that graphene bending, interaction with substrate, and defects can effectively relax this selection rule[50-52].

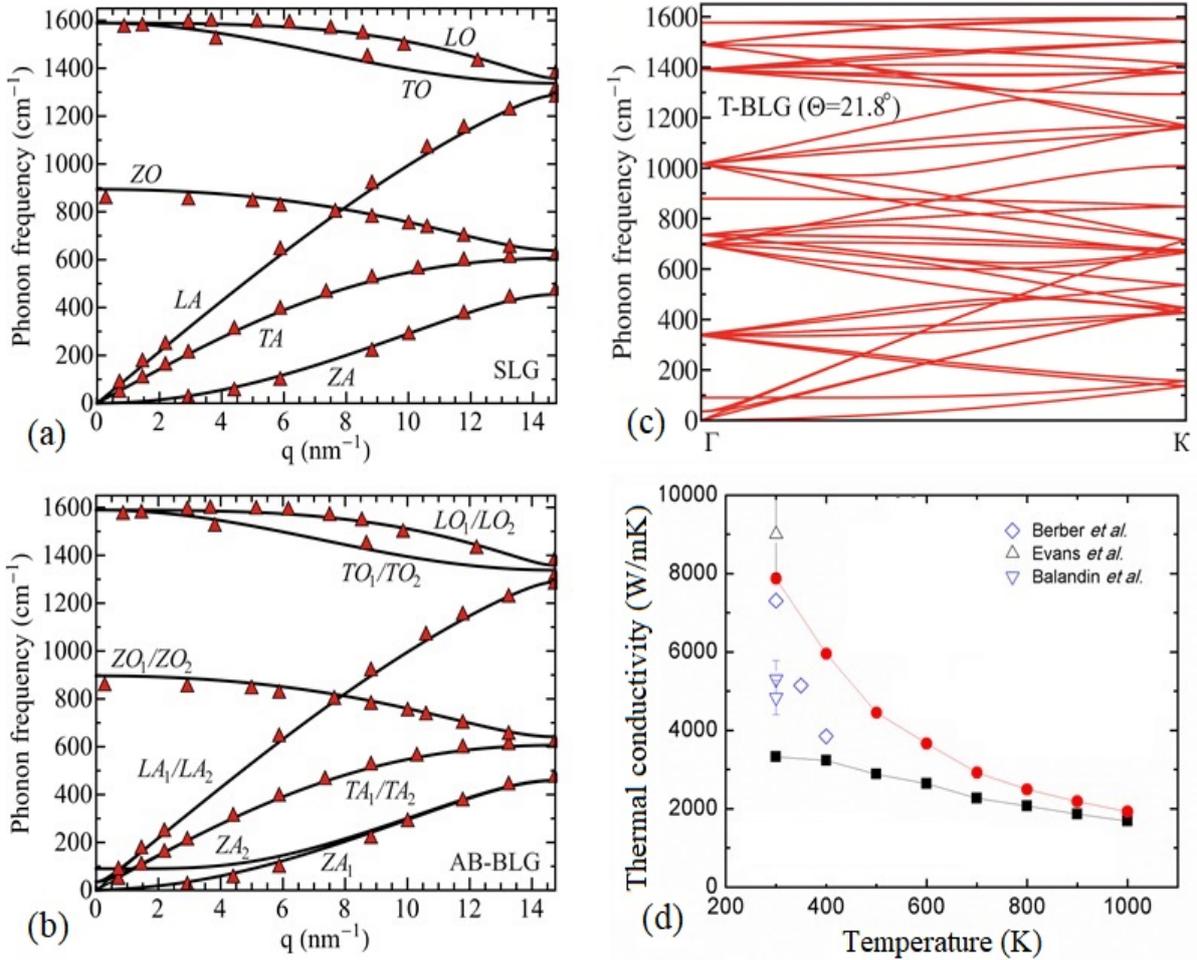

**Figure 2:** Phonon dispersions in (a) single layer graphene and (b) AB-stacked bilayer graphene, plotted along Γ-K direction of Brillouin zone. (c) Phonon dispersion in twisted bilayer graphene with the twisting angle 21.8º. (d) Thermal conductivity of suspended single-layer graphene as a function of temperature calculated for the phonon number obeying the Bose-Einstein and classical statistics. The data points in (d) shown by diamonds and triangles are from Ref. [50]. The figures are adapted from Refs. [15, 20, 50].

## IV. Thermal Properties of Graphene-Enhanced Composites

While graphene has the highest intrinsic thermal conductivity, FLG is the most promising material for practical applications in thermal interface materials (TIMs). This assessment is based on the observations that (i) FLG possesses a high thermal conductivity, in the range from 500 W/mK to 2000 W/mK, depending on the quality; (iii) FLG has a larger cross-section area than graphene to conduct heat though; (iv) the thermal conductivity of FLG degrades less upon exposure to matrix material in the composites; (v) FLG retains mechanical flexibility required for thermal coupling to the matrix material; and (vi) FLG with variable thickness can be mass-produced at low cost. In the context of thermal management, the term "graphene" typically refers to FLG with a thickness range from a few atomic planes to tens of nanometers and lateral dimensions of a few microns. It is imperative that the lateral dimensions be greater than a micrometer (μm) to be above the grey phonon MFP in graphene. In this sense, graphene – FLG fillers used in thermal composites are different from graphite nano-platelets, characterized by smaller lateral dimensions and aspect ratios, and from milled graphite fillers with hundreds of nanometers or micrometer thicknesses. Much thicker graphite fillers do not have the flexibility of FLG and, as a result, do not couple well to the matrix. The first studies[21] of graphene composites found that even small loading fractions of graphene fillers can increase the thermal conductivity of composites by up to a factor of ×25 (see Figure 3). These results have been independently confirmed[24-25]. A recent report[26] of composites with high loading of graphene revealed clear signatures of thermal percolation, which allowed for the achievement of the thermal conductivity of ~12 W/mK, exceeding that of commercial TIMs. Orientation of FLG fillers further increases thermal conductivity of composites[53-54].

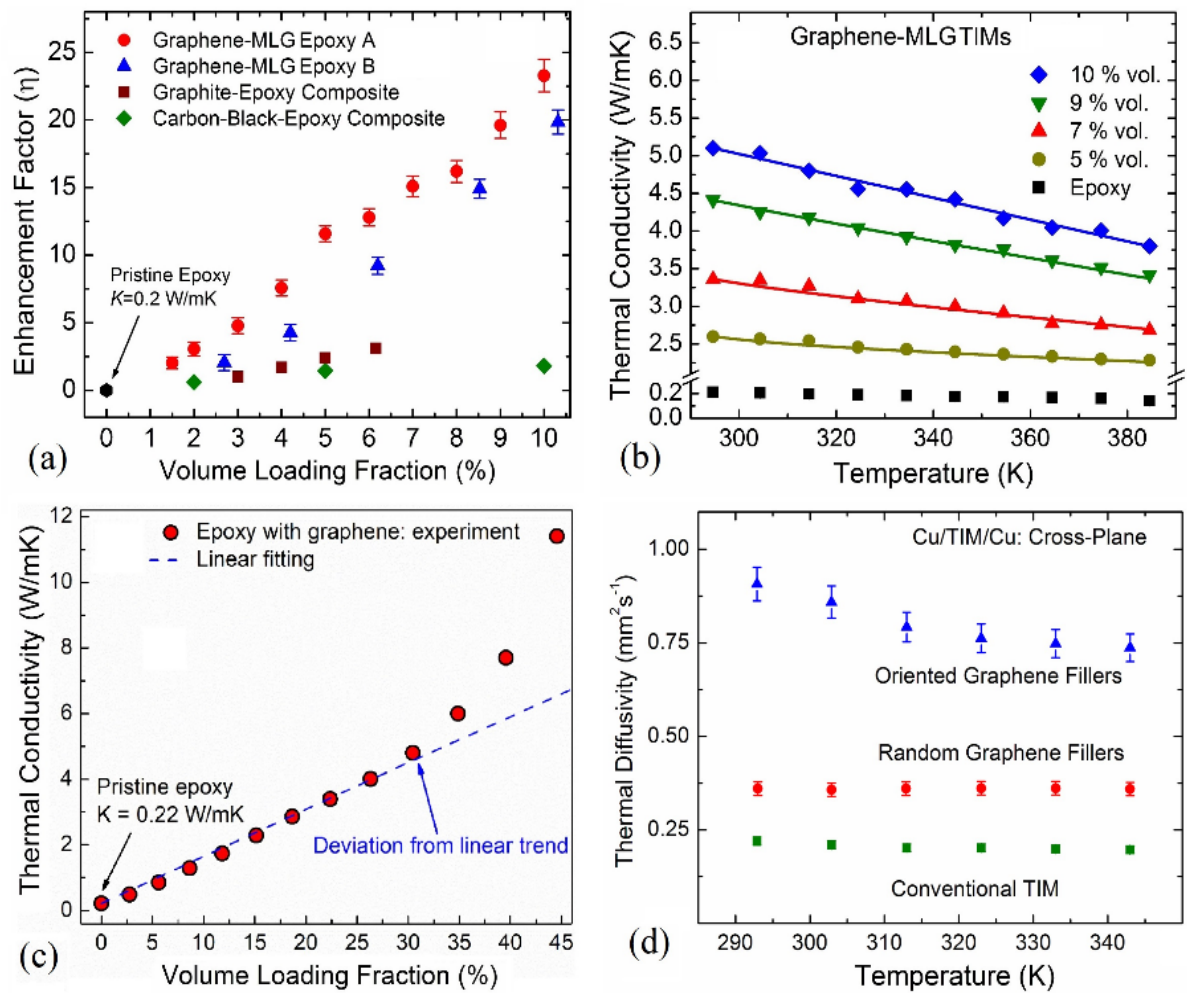

**Figure 3:** Thermal conductivity of the graphene composites. (a) Thermal conductivity enhancement factor as a function of the filler loading fraction. (b) Temperature dependent thermal conductivity of graphene – epoxy composites for different graphene loadings. (c) Thermal conductivity of the epoxy composites with graphene at high filler loading. Thermal conductivity depends linear on the loading until a threshold value, and then becomes super-linear, indicating the onset of the thermal percolation transport regime. (d) Apparent thermal diffusivity of Cu-TIM-Cu sandwiches with graphene-enhanced TIMs and that of a reference TIM without graphene. Note the importance of graphene filler orientation. The figures are adapted from Ref. [21, 23, 26, 53].

**Acknowledgements:** I thank my former PhD students at UC Riverside who contributed to graphene phononics research: Dr. Irene Calizo, Dr. Suchismita Ghosh, Dr. Samia Subrina, Dr. Khan Shahil, Dr. Vivek Goyal, Dr. Guanxiong Liu, Dr. Zhong Yan, Dr. Hoda Malekpour, Dr. Jackie Renteria, and Dr. Mohammed Saadah. Special thanks go to the current group members, particularly Dr. Fariborz Kargar and PhD students Zahra Barani, Jacob Lewis, and Sahar Naghibi for their research of graphene's applications in thermal interface materials. I am indebted to the late Professor Evgeni Pokatilov and Professor Denis Nika for their valuable contributions to the theory of phonon transport in graphene. My group's phononics research is presently supported by NSF, DARPA, and DOE via EFRC SHINES at UCR.